\documentclass[times,a4paper,12pt]{elsarticle}

\usepackage{amsmath}
\usepackage{graphicx}
\usepackage{hyperref}
\usepackage{xcolor}

\newcommand{\de}[1]{\,{\mathrm d}#1}

\newcommand{\var}{\delta}

\newcommand{\mtrx}[1]{\mathsf{#1}} 
\newcommand{\trn}{{\sf ^T}}        

\newcommand{\Normal}{N_x}
\newcommand{\Shearz}{V_z}
\newcommand{\Bendy}{M_y}

\newcommand{\rot}{\varphi}

\newcommand{\strain}{\varepsilon}
\newcommand{\sstrain}{\gamma}

\newcommand{\stress}{\sigma}
\newcommand{\sstress}{\tau}

\newcommand{\del}[2]{\mbox{$\displaystyle\frac{#1}{#2}$}}
\newcommand{\ppd}[2]{\del{\partial {#1}}{\partial{#2}}}

\newcommand{\der}[2]{\del{\de #1}{\de #2}}

\newcommand{\layer}[1]{^{(#1)}}
\newcommand{\el}{_e}
\newcommand{\ele}[1]{_{e,#1}}

\hypersetup{%
  pdfauthor={A. Zemanova, J. Zeman, M. Sejnoha},%
  pdftitle={Simple Numerical Model of Laminated Glass Beams},%
  pdfkeywords={laminated glass beams, finite element method, Lagrange multipliers},%
  colorlinks%
}

\journal{arxiv}

\begin{document}

\begin{frontmatter}

\title{Simple Numerical Model of Laminated Glass Beams} 

\author[mech]{Alena Zemanov\'{a}}
\ead{zemanova.alena@gmail.com}
\author[mech]{Jan Zeman\corref{cor}}
\ead{zemanj@cml.fsv.cvut.cz}
\ead[url]{http://mech.fsv.cvut.cz/~zemanj} 
\author[mech,cideas]{Michal \v{S}ejnoha}
\ead{sejnom@fsv.cvut.cz}

\cortext[cor]{Corresponding author. Tel.:~+420-2-2435-4482; fax~+420-2-2431-0775}
\address[mech]{%
Department of Mechanics, Faculty of Civil Engineering, Czech Technical
University in Prague, Th\' akurova 7, 166 29 Prague 6, Czech Republic}

\address[cideas]{%
Centre for Integrated Design of Advances Structures, Th\' akurova 7,\\
166 29 Prague~6, Czech Republic}

\begin{abstract}
This contribution presents a simple Finite Element model aimed at
efficient simulation of layered glass units. The adopted approach is
based on considering independent kinematics of each layer, tied
together via Lagrange multipliers. Validation and verification of the
resulting model against independent data demonstrate its accuracy,
showing its potential for generalization towards more complex
problems.
\end{abstract}


\end{frontmatter}

\section{Introduction}\label{sec:intro}
%
The most frequently used transparent material in the building
envelopes is glass. It is a fragile material, which fails in a brittle
manner. This is the reason for using safety glasses in a situation
when there is a possibility of human impact or where the glass could
fall if shattered.

Laminated glass is a multi-layer material produced by bonding two or
more layers of glass together with a plastic interlayer, typically
made of polyvinyl butyral (PVB). The interlayer keeps the layers of
glass bonded even when broken, and its high strength prevents the
glass from breaking up into large sharp pieces. This produces a
characteristic "spider web" cracking pattern when the impact is not
powerful enough to completely pierce the glass. Multiple laminae and
thicker glass decrease stress level,
  thereby increasing the load-carrying capacity of a structural
member, too.

The focus of this study is on the establishing a simple and versatile
framework for the analysis of mechanical behavior of laminated glass
units. To keep the discussion compact, we restrict our attention to
the linearly elastic response of layered glass beams in the small
strain regime. The rest of the paper is organized as follows. Methods
of analysis of laminated glass beams are introduced in
Section~\ref{sec:method}, together with a brief characterization of
the proposed numerical model. The principles of the method are
described in detail in Sections~\ref{sec:formulation}
and~\ref{sec:discretization}. In particular, the mechanical
formulation of the model is shown in Section~\ref{sec:formulation}. In
the next section, the Finite Element discretization is presented. In
Section~\ref{sec:results}, the proposed numerical technique is
verified and validated against a reference analytical solution and
publicly available experimental data. Finally, Section~\ref{sec:concl}
concludes the paper and discusses future extensions of the method.

\section{Brief overview of available methods}\label{sec:method}
%
The most frequent approach to the analysis of glass structural
elements was, for a long time, based on empirical knowledge. Such
relations are sufficient for the design of traditional windows
glasses. In modern architecture there has been a steadily growing
demand on the use of transparent materials for large external walls
and roof systems in the recent decades. Therefore, the detailed
analysis of layered glass units is becoming increasingly important in
order to ensure a reliable and efficient design.

In general, the complex behavior of laminated glass can be considered
as an intermediate state of two limiting
cases~\cite{Vallabhan:1987:SLG}. In the first case, the structure is
treated as an assembly of two independent glass plates without any
interlayer~(the lower bound on stiffness and strength of a member),
while in the second case, corresponding to the upper estimate of
strength and stiffness, the glass unit is modeled as a monolithic
glass (one glass plate with thickness equal to the total thickness of
the glass plates).  Both elementary cases, however, fail to correctly
capture complex interaction among individual layers, leading to
non-optimal layer thickness designs. Therefore, several alternative
approaches to the analysis of layered glass structures have been
proposed in the literature. These methods can be categorized into
three basic groups:

\begin{itemize}
\item methods calibrated with respect to experimental
  measurements~\cite{Norville:1998:BSLG},
\item analytical
  approaches~\cite{Vallabhan:1993:ALG,Asik:2003:LGP,Asik:2005:MMB},
\item numerical models typically based on detailed Finite Element
  simulations~\cite{Duser:1999:AGBL,Ivanov:2006:AMO}.
\end{itemize} 

Applicability of analytical approaches to practical (usually
large-scale) structures is far from being straightforward. In
particular, the closed-form solution of the resulting equations is
possible only for very specific boundary conditions and therefore have
to be solved by an appropriate numerical method. Moreover, the
analytical approaches are rather difficult to be generalized to beams
with multiple layers. Therefore, it appears to be advantageous to
directly formulate the problem in the discretized form, typically
based on the Finite Element Method~(FEM). Nevertheless, we would like
to avoid fully resolved two- or three-dimensional simulations,
cf.~\cite{Duser:1999:AGBL,Ivanov:2006:AMO}, which lead to
unnecessarily expensive calculations.

In this paper, we propose a simple FEM model inspired by a specific
class of refined plate
theories~\cite{Mau:1973:RLP,Sejnoha:1996:MMU,Matous:1998:EBO}. In this
framework, each layer is treated as the Timoshenko beam with
independent kinematics. Interaction between individual layers is
captured by the Lagrange multipliers (with a physical meaning of shear
stresses), which result from the conditions of inter-layer
displacements compatibility. Such a refined approach circumvents the
limitation of similar models available in typical commercial FEM
systems, which employ a single set of kinematic variables and average
the mechanical response through the thickness of the beam,
e.g.~\cite{Bathe:1996:FEM}. Unlike the proposed approach, the
averaging operation is too coarse to correctly represent the
inter-layer interactions, see Section~\ref{sec:results} for a concrete
example.

\section{Mechanical model of laminated beams}\label{sec:formulation}
\begin{figure}[ht]
\centerline{%
 \includegraphics[height=70mm]{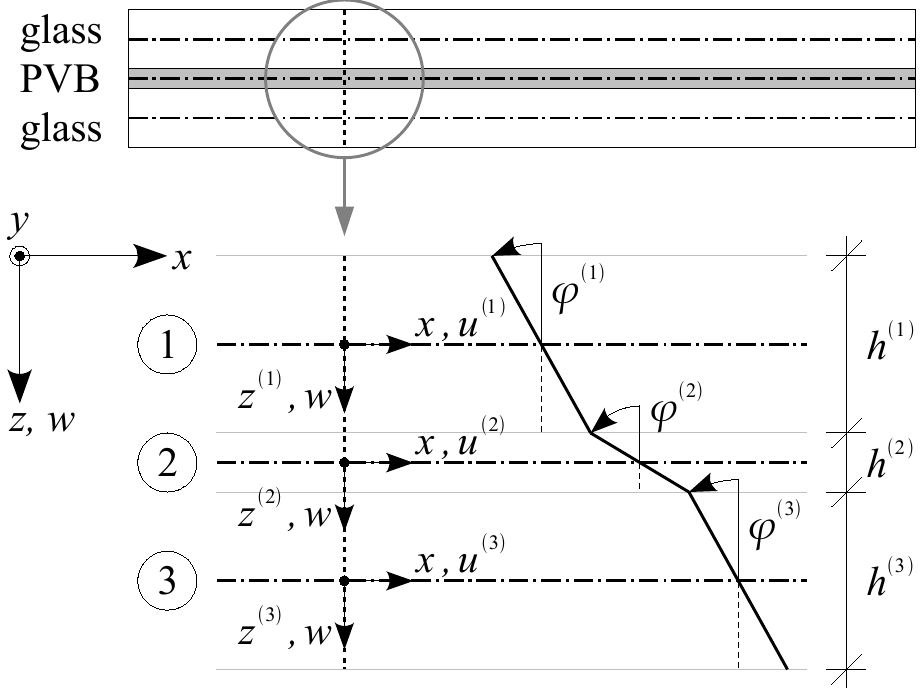} 
}%
\caption{Kinematics of laminated beam}
\label{fig:lam_beam}
\end{figure}

%
As illustrated in Figure~\ref{fig:lam_beam}, laminated glasses consist
mostly of three layers. A local coordinate system is attached to each
layer to allow for an efficient treatment of independent
kinematics. In the following text, a quantity $a$ expressed in the
local coordinate system associated with the $i$-th layer is denoted as
$a\layer{i}$, whereas a variable without an index represents a
globally defined quantity, cf. Figure~\ref{fig:lam_beam}.

Each layer is modeled using the Timoshenko beam theory supplemented
with membrane effects. Hence, the following kinematic assumptions
are adopted
\begin{itemize}
\item the cross sections remain planar but not necessarily
  perpendicular to the deformed beam axis,
\item vertical displacement does not vary along the height of the beam
  (when compared to the magnitude of the displacement).
\end{itemize}
Under these assumptions, the non-zero displacement components in each layer are parametrized as:
\begin{eqnarray*}
u\layer{i}(x,z\layer{i}) &=& u\layer{i}(x,0) + \rot\layer{i}(x) z\layer{i}, \\ w\layer{i}(x,z\layer{i}) &=& w(x),
\end{eqnarray*}
where $i=1,2,3$ and $z\layer{i}$ is measured in the local coordinate
system from the middle plane of the $i$-th layer. The inter-layer
interaction is ensured via the continuity conditions specified on
interfaces between layers in the form~($i=1,2$)
\begin{eqnarray}
u\layer{i}(x,\frac{h\layer{i}}{2}) - u\layer{i+1}(x,-\frac{h\layer{i+1}}{2})= 0.
\label{eq:continuity}
\end{eqnarray}
The strain field in the $i$-th layer follows from the strain-displacement relations~\cite{Bittnar:1996:NMM,Bathe:1996:FEM}
\begin{eqnarray}
\strain\layer{i}_{x} (x,z\layer{i}) &=& \ppd{u\layer{i}}{x}(x,z\layer{i}) = \der{u\layer{i}}{x}(x,0) + \der{\rot\layer{i}}{x}(x) z\layer{i}, 
\nonumber \\
\sstrain\layer{i}_{xz} (x) &=& \ppd{u\layer{i}}{z\layer{i}}(x,z\layer{i}) + \ppd{w}{x}(x) = \rot\layer{i}(x) + \der{w}{x}(x),
\label{eq:shear_strain_def}
\end{eqnarray}
which, when combined with the constitutive equations of each layer
expressed in terms of Young's modulus $E$ and the shear modulus $G$:
\begin{eqnarray*}
\stress\layer{i}_{x} (x,z\layer{i}) = E\layer{i} \strain\layer{i}_{x}(x,z\layer{i}) &\mbox{and}&
\sstress\layer{i}_{xz} (x) = G\layer{i} \sstrain\layer{i}_{xz}(x),
\end{eqnarray*}
yield the expressions for the internal forces as
\begin{eqnarray*}
\Normal\layer{i} (x) &=& E\layer{i} A\layer{i} \der{u\layer{i}}{x}(x,0), \\
\Shearz\layer{i} (x) &=& k G\layer{i} A\layer{i} \left(\rot\layer{i}(x) + \der{w}{x}(x)\right), \\
\Bendy\layer{i}  (x) &=& E\layer{i} I\layer{i} \der{\rot\layer{i}}{x}(x),
\end{eqnarray*}
where $b$ and $h\layer{i}$ are the width and height of the $i$-th
layer, recall Figure~\ref{fig:lam_beam}, and $k =\frac{5}{6}$,
$A\layer{i}=b h\layer{i}$ a $I\layer{i}=\frac{1}{12} b (h\layer{i})^3$
stand for the shear correction factor, the cross-section area and the
moment of inertia of the $i$-th layer, respectively.

To proceed, consider the weak form of the governing equations, written
for the $i$-th layer~(the subscripts $\bullet_x$ and $\bullet_z$
related to internal forces and kinematics-related quantities are
omitted in the sequel for the sake of brevity)
\begin{eqnarray*}
\int\limits_{0}^{L} \der{}{x}\left(  \var u\layer{i}(x) \right) E\layer{i} A\layer{i} \der{}{x}\left( u\layer{i}(x) \right) \de x
=
\int\limits_{0}^{L} \var u\layer{i} (x) \bar{f}\layer{i}_x (x) \de x + \left[\var u (x) \bar{N}\layer{i}(x) \right]_{0}^{L},
\\
\int\limits_{0}^{L} \der{}{x}\left(  \var w (x) \right) k G\layer{i} A\layer{i} \sstrain\layer{i} (x) \de x
=
\int\limits_{0}^{L} \var w (x) \bar{f}\layer{i}_z (x) \de x + \left[\var w (x) \bar{V}\layer{i}(x) \right]_{0}^{L},
\\
\int\limits_{0}^{L} \der{}{x}\left(  \var \rot\layer{i}(x) \right) E\layer{i} I\layer{i} \der{}{x}\left( \rot\layer{i}(x) \right) \de x
=
\left[\var \rot\layer{i}(x) \bar{M}\layer{i}(x) \right]_{0}^{L},
\\
\int\limits_{0}^{L} 
\var \sstrain\layer{i} (x) k G\layer{i} A\layer{i}
\left[\sstrain\layer{i} (x) - \rot\layer{i}(x) - \der{}{x}\left(  w (x) \right) \right]\de x
= 0.
\end{eqnarray*}
to be satisfied for arbitrary admissible test fields $\var
u\layer{i}$, $\var \rot\layer{i}$ and $\var w$. In particular, the
first three equations correspond to equilibrium conditions written for
normal and shear forces and bending moments, respectively. The last
identity enforces the geometrical relation~\eqref{eq:shear_strain_def}
in the integral form, thereby allowing to treat the shear strain as an
independent field in the discretization procedure discussed
next. Further note that the continuity
conditions~\eqref{eq:continuity} will be introduced directly into the
discretized formulation, as explained in the following Section.

\section{Finite element discretization}\label{sec:discretization}
\begin{figure}[ht]
\centerline{%
 \includegraphics[height=55mm]{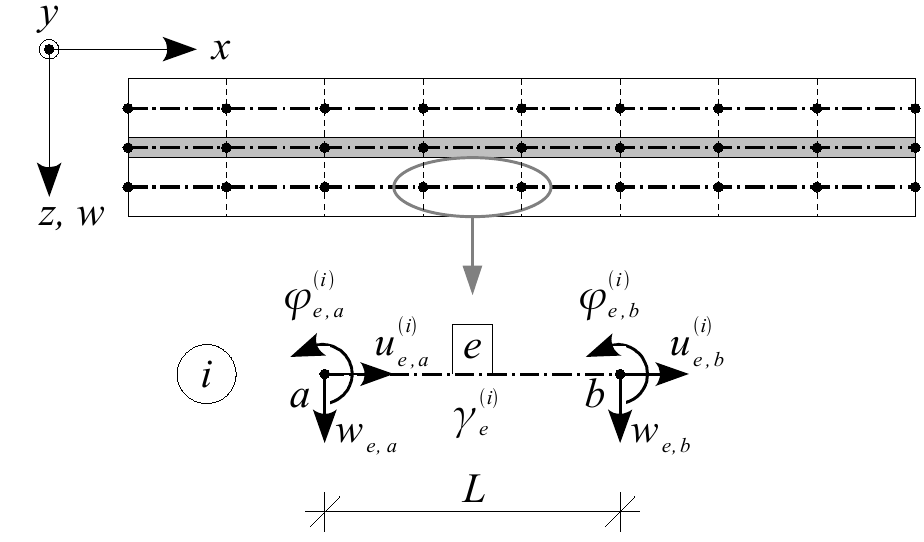} 
}%
\caption{Finite element discretization of the $i$-th layer}
\label{fig:fin_elem}
\end{figure}

%
To keep the discretization procedure transparent, it is assumed that
each layer of the laminated beam is divided into identical number of
elements, leading to the discretization scheme illustrated in
Figure~\ref{fig:fin_elem}.

Following the standard conforming Finite Element machinery,
e.g.~\cite{Bittnar:1996:NMM,Bathe:1996:FEM}, we express the searched and test
displacement fields at the element level in the form
\begin{eqnarray*}
u\layer{i}\el(x) \approx \mtrx{N}\layer{i}\ele{u}(x) \mtrx{r}\layer{i}\ele{u}, &&
\var u\layer{i}\el(x) \approx \mtrx{N}\layer{i}\ele{u}(x) \var \mtrx{r}\layer{i}\ele{u}, \\
w\el(x) \approx \mtrx{N}\ele{w}(x) \mtrx{r}\ele{w}, &&
\var w\el(x) \approx \mtrx{N}\ele{w}(x) \var \mtrx{r}\ele{w}, \\
\rot \layer{i}\el(x) \approx \mtrx{N}\layer{i}\ele{\rot}(x) \mtrx{r}\layer{i}\ele{\rot},&&
\var \rot \layer{i}\el(x) \approx \mtrx{N}\layer{i}\ele{\rot}(x) \var \mtrx{r}\layer{i}\ele{\rot}, \\
\sstrain \layer{i}\el(x) \approx \mtrx{N}\layer{i}\ele{\sstrain}(x) \mtrx{r}\layer{i}\ele{\sstrain},&&
\var \sstrain \layer{i}\el(x) \approx \mtrx{N}\layer{i}\ele{\sstrain}(x) \var \mtrx{r}\layer{i}\ele{\sstrain},
\end{eqnarray*}
where $e$ is used to denote the element number, $\bullet\el$ and
$\var\bullet\el$ denote a relevant searched and test field restricted
to the $e$-th element, $\mtrx{N}\ele{\bullet}\layer{i}$ is the
associated matrix of basis functions and
$\mtrx{r}\layer{i}\ele{\bullet}$ the matrix of nodal unknowns. In the
actual implementation, the fields $u\layer{i}$, $w\el$ and $\rot
\layer{i}\el$, as well as the corresponding test quantities, are
assumed to be piecewise linear. To obtain a locking-free element, the
shear strain $\sstrain \layer{i}\el$ is taken as constant and is
eliminated using the static condensation,
see~\cite{Bittnar:1996:NMM,Bathe:1996:FEM} for additional details.

To simplify the further treatment, we consider the following
partitioning of the stiffness matrix $\mtrx{K}$ and the right hand
side matrix $\mtrx{R}$ related to the $e$-th element and the $i$-th
layer after the static condensation:
\begin{equation*}
\left[
\begin{array}{cc}
\mtrx{ K }\layer{i}\el & \mtrx{ K }\layer{i}_{ew} \\
\mtrx{ K }\layer{i}_{we} & \mtrx{ K }\layer{i}_{w} 
\end{array}
\right]
\left[
\begin{array}{c}
\mtrx{r}\layer{i}\el \\
\mtrx{r}\ele{w} 
\end{array}
\right]
=
\left[
\begin{array}{c}
\mtrx{R}\layer{i}\el \\
\mtrx{R}\layer{i}\ele{w} 
\end{array}
\right],
\end{equation*}
where $\mtrx{ K }\layer{i}_{ew} = \left( \mtrx{ K }\layer{i}_{we} \right) \trn$ and
\begin{eqnarray*}
\mtrx{r}\layer{i}\el = \left[ u\layer{i}\ele{a}, u\layer{i}\ele{b}, \rot\layer{i}\ele{a}, \rot\layer{i}\ele{b} \right]\trn, && 
\mtrx{r}\ele{w} = \left[ w\ele{a}, w\ele{b} \right]\trn.
\end{eqnarray*}
Considering all three layers in Figure~\ref{fig:fin_elem} together
gives the resulting system of governing equations in the form
\begin{equation*}
\left[
\begin{array}{ccccc}
\mtrx{ K }\layer{1}\el & \mtrx{ 0 } & \mtrx{ 0 } & \mtrx{ K }\layer{1}_{ew} & \\
\mtrx{ 0 } & \mtrx{ K }\layer{2}\el & \mtrx{ 0 } & \mtrx{ K }\layer{2}_{ew} & \mtrx{ E }\el \trn \\
\mtrx{ 0 } & \mtrx{ 0 } & \mtrx{ K }\layer{3}\el & \mtrx{ K }\layer{3}_{ew} & \\
\mtrx{ K }\layer{1}_{we} & \mtrx{ K }\layer{2}_{we} & \mtrx{ K }\layer{3}_{we} & 
\mtrx{ K }\layer{1}_{w} + \mtrx{ K }\layer{2}_{w} + \mtrx{ K }\layer{3}_{w} & \mtrx{ 0 } \\
& \mtrx{ E }\el & & \mtrx{ 0 } & \mtrx{ 0 }
\end{array}
\right]
\left[
\begin{array}{c}
\mtrx{r}\layer{1}\el \\
\mtrx{r}\layer{2}\el \\
\mtrx{r}\layer{3}\el \\
\mtrx{r}\ele{w} \\
\mtrx{\lambda}_{(4 \times 1)}
\end{array}
\right]
=
\left[
\begin{array}{c}
\mtrx{R}\layer{1}\el \\
\mtrx{R}\layer{2}\el \\
\mtrx{R}\layer{3}\el \\
\mtrx{R}\layer{1}\ele{w} + \mtrx{R}\layer{2}\ele{w} + \mtrx{R}\layer{3}\ele{w} \\
\mtrx{ 0 }
\end{array}
\right],
\end{equation*}
where the matrix $\mtrx{\lambda}$ stores the nodal values of the
Lagrange multipliers, associated with the compatibility
constraint~\eqref{eq:continuity}, and the matrix
\begin{equation*}
\mtrx{ E }\el 
=
\left[
\begin{array}{cccccccccccc}
1 & 0 & \frac{h\layer{1}}{2} & 0 & -1 & 0 & \frac{h\layer{2}}{2} & 0 & 0 & 0 & 0 & 0 \\
0 & 1 & 0 & \frac{h\layer{1}}{2} & 0 & -1 & 0 & \frac{h\layer{2}}{2} & 0 & 0 & 0 & 0 \\ 
0 & 0 & 0 & 0 & 1 & 0 & \frac{h\layer{2}}{2} & 0 & -1 & 0 & \frac{h\layer{3}}{2} & 0 \\
0 & 0 & 0 & 0 & 0 & 1 & 0 & \frac{h\layer{2}}{2} & 0 & -1 & 0 & \frac{h\layer{3}}{2}
\end{array}
\right]
\end{equation*}
implements the tying conditions.

\section{Verification and validation of numerical model}\label{sec:results}
%
\begin{figure}[ht]
\centerline{%
 \includegraphics[height=45mm]{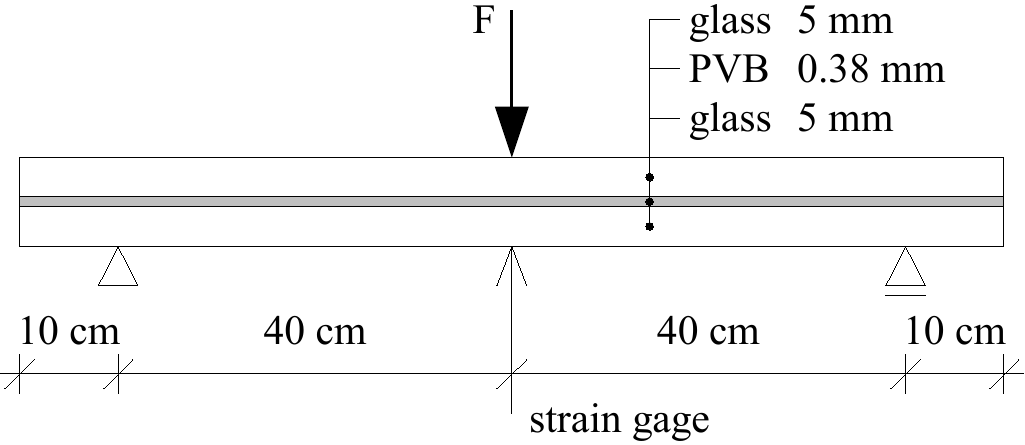} 
}%
\caption{Three point bending setup for simply supported beam}
\label{fig:bend_test}
\end{figure}
To verify and validate the performance of the present approach, the
previously described FEM model was implemented using
MATLAB\textregistered~system and compared against predictions of an
analytical model and experimental data for a three-point bending test
on a simply supported laminated glass beam presented
in~\cite{Asik:2005:MMB}, see also Figure~\ref{fig:bend_test}. The
width of the beam is $b=0.1$~m and material data of individual
components of the structure are available in
Table~\ref{t:material_data}. The glass modulus of
  elasticity is slightly lower than the conventional values of
  $70$--$73$~GPa reported in the literature and is specific for the
  material employed in the experiment. Moreover, as the PVB layer
  shows viscoelastic and temperature-dependent behavior, the modulus
  of elasticity corresponds to an effective secant value related to
  load duration of $60$~s and temperature of $22^\circ$~C.
\begin{table}[ht]
\caption{Material data} 
\label{t:material_data}
\centerline{%
\begin{tabular}{|l|cc|}
\hline
& \emph{Glass} & \emph{PVB layer} \\
\hline
Young's modulus, $E$   & $64.5$~GPa & $1.287$~MPa \\
Poisson's ratio, $\nu$ & $0.23$ & $0.4$ \\
\hline
\end{tabular}
}
\end{table}

Table~\ref{t:porovnani3} summarizes values of the mid-span deflection
for a representative load level determined by FE-based discretization
using 60 elements (30 when symmetry of the problem is
  exploited) and the corresponding experimental
values. Note that the discretization is sufficient to
  achieve three-digit accuracy in the mid-span deflection. In
addition to the results obtained by an analytical method proposed by
Asik and Tezcan in~\cite{Asik:2005:MMB}, the results of the analysis
using ADINA\textregistered~system and the elementary lower and upper
bounds are included. In particular, the ADINA\textregistered~model is
based on the classical laminate theory, cf.~\cite{Bathe:1996:FEM},
whereas the two simplified approaches assume zero or infinite
compliance of the interlayer, recall also discussion in
Section~\ref{sec:method}. In the following discussion,
  e.g. $\eta^{num}_{exp}$ denotes the relative error between the
  numerical prediction and reference experimental value, while
  e.g. $\eta_{an}$ is used for the error of analytical solution when
  compared to candidate approaches.  Clearly, the results of the last
three methods differ substantially from experimental data as well as
the analytical results. The proposed numerical model, on the other
hand, shows a response almost identical to the analytical method,
which deviates from experimental measurement by less then $6\%$. Such
accuracy can be considered as sufficient from the practical point of
view.

\begin{table}[ht]

\caption{Comparison of results for a simply supported beam (load 50
  N)}
\label{t:porovnani3}

\centerline{%
\begin{tabular}{|llccc|}
\hline
Model	& &	Central deflection [mm]	& $\eta_{exp}$	& $\eta_{an}$ \\
\hline
\multicolumn{5}{|l|}{Laminated glass beam: thickness [mm] 5/0.38/5 (glass/PVB/glass)} \\
&	Experiment &	1.27	& -	& -5.2$\%$ \\
&	Analytical model	& 1.34	& 5.5$\%$	& - \\
&\bfseries	Numerical model	& \bfseries 1.34	& \bfseries 5.5$\%$	& \bfseries0.0$\%$ \\
&	ADINA \textregistered (Multi-layered shell)	& 0.89	& -30.2$\%$ &	-33.8$\%$ \\
\hline
\multicolumn{5}{|l|}{Monolithic glass beam: thickness [mm] 10 (glass+glass)} \\
&	Analytical model	& 0.99	& -21.8$\%$	& -25.9$\%$ \\
\hline
\multicolumn{5}{|l|}{Two independent glass beams: thickness [mm] 5/5 (without any interlayer)} \\
&	Analytical model	& 3.97	& 212.6$\%$	& 196.2$\%$ \\
\hline
\end{tabular}
}
\end{table}

To further confirm predictive capacities of the proposed numerical
scheme, a response corresponding to a proportionally increasing load
was investigated. The results appear in Tables~\ref{t:porovnani}
and~\ref{t:porovnani2}. Again, the method seems to be sufficiently
accurate in the investigated range of loads when considering the
values of deflections as well as values of local stresses and strains.

\begin{table}[ht]
\caption{Comparison of deflections for a simply supported beam} 
\label{t:porovnani}
\centerline{%
\begin{tabular}{|c|cccccc|}
\hline
Load [N] & \multicolumn{6}{c|}{Central deflection [mm]} \\
& $w_{exp}$ & $w_{an}$ & $\eta^{an}_{exp} $ [\%] & $w_{num}$ & $\eta^{num}_{exp}$ [\%] & $\eta^{num}_{an} $ [\%] \\
\hline
50&     1.27&	1.34&	 5.51&	1.34&	 5.51&	 0.00\\
100&	2.55&	2.69&	 5.49&	2.68&	 5.10&	-0.37\\
150&	4.12&	4.03&	-2.18&	4.02&	-2.43&	-0.25\\
200&	5.57&	5.38&	-3.41&	5.36&	-3.77&	-0.37\\
\hline
\end{tabular}
}
\end{table}
\begin{table}[ht]
\caption{Comparison of stresses and strains for a simply supported
  beam}
\label{t:porovnani2}
\centerline{%
\begin{tabular}{|c|ccc|ccc|}
\hline
Load [N] & \multicolumn{3}{c|}{Maximum strain [$\times 10^{-6}$]} & \multicolumn{3}{c|}{Maximum stress [MPa]}  \\
& $\epsilon_{an}$ & $\epsilon_{num}$ & $\eta^{num}_{an} $ [\%]
& $\sigma_{an}$ & $\sigma_{num}$ & $\eta^{num}_{an} $ [\%] \\
\hline
50 &	112&	114&	1.79&	 7.23&   7.34&  1.52\\
100&	224&	228&	1.79&	14.45&	14.68&  1.59\\
150&	336&	341&	1.49&	21.68&	22.02&	1.57\\
200&	448&	455&	1.56&    28.9&  29.36&	1.59\\
\hline
\end{tabular}
}
\end{table}

\section{Conclusions}\label{sec:concl}
%
As shown by the presented results, the proposed numerical method is
well-suited for the modeling of laminated glass beams, mainly because
of its low computational cost and accurate representation of the
structural member behavior. Future improvements of the model will
consider large deflections and the time-dependent response of the
interlayer and will be reported separately.

\section*{Acknowledgments}
%
The support provided by the GA\v CR grant No. 106/07/1244 is gratefully acknowledged.

\bibliographystyle{elsarticle-num} 
\bibliography{liter}

\end{document}